\documentclass[10pt, a4paper]{article}
\usepackage{lrec2000}

\title{Enriching a Text by Semantic Disambiguation for Information Extraction}

\name{Bernard Jacquemin, Caroline Brun and Claude Roux}

\address{Xerox Research Centre Europe\\
         6, chemin de Maupertuis, 38\,240 Meylan, France\\ 
         \{Bernard.Jacquemin,Caroline.Brun,Claude.Roux\}@xrce.xerox.com}

\abstract{External linguistic resources have been used for a very long time in information extraction. These methods enrich a document with data that are semantically equivalent, in order to improve recall. For instance, some of these methods use synonym dictionaries. These dictionaries enrich a sentence with words that have a similar meaning. However, these methods present some serious drawbacks, since words are usually synonyms only in restricted contexts. The method we propose here consists of using word sense disambiguation rules (WSD) to restrict the selection of synonyms to only these that match a specific syntactico-semantic context. We show how WSD rules are built and how information extraction techniques can benefit from the application of these rules.}

\begin{document}

\maketitleabstract

\section{Introduction}

In today's world, the society of communications is gaining in importance every day. The amount of electronic documents -- mainly by Internet, but not only -- grows more and more. With this increase, no one is able to read, classify and structure those documents so that the requested information can be reached when it is needed. Therefore we need tools that reach a shallow understanding of the content of these texts to help us to select the requested data.

The process of understanding a document consists in identifying the concepts of the document that correspond to requested information. This operation can be performed with linguistic methods that permit the extraction of various components related to the data that are requested.

Since the beginning of the '90s, several research projects in information extraction from electronic text have been using linguistic tools and resources to identify relevant elements for a request. The first ones, based on domain-specific extraction patterns, use hand-crafted pattern dictionaries (CIRCUS \cite{Lehnert90}). But systems were quickly designed to build extraction pattern dictionaries automatically. Among these systems, AutoSlog \cite{Riloff93,RiloffLorenzen99} builds extraction pattern dictionaries for CIRCUS. CRYSTAL \cite{SoderlandAl95} creates extraction patterns lists for BADGER, the successor of CIRCUS. These learners use hand-tagged specific corpora to identify structures containing the relevant information. The syntactic structure used by CRYSTAL is more subtle than the one used by AutoSlog. CRYSTAL is able to make the most of semantic classes. WHISK \cite{Soderland99} is one of the most recent information extraction system. WHISK has been designed to learn which data to extract from structured, semi-structured and free text~\footnote{We use the term ``structured text'' to refer to what the database community calls semi-structured text; ``semi-structured text'' is ungrammatical and often telegraphic text that does not follow any rigid format; ``free text'' is simply grammatical text \cite{Soderland99}.}. A parser and a semantic tagger have been implemented for free text. This system is the only one to process all of these three categories of text.

These methodologies need domain-specific pattern dictionaries that must be built for each different kind of information. However, none of these methods can be directly applied to generic information. Thus we decide to bypass these two obstacles: our approach is based on the utilization of an existing electronic dictionary, in order to expand the data in a document to equivalent forms extracted from that dictionary.

Our method deals with the identification of semantic contents in documents through a lexical, syntactic and semantic analysis. It then becomes possible to enrich words and multi-word expressions in a document with synonyms, synonymous expressions, semantic information etc.\ extracted from the dictionary. 

\section{Problems and Prospects}

As for a lot of methodologies developed for natural language processing, the results of a method of information extraction are evaluated by two measures: precision and recall. Precision is the ratio of correctly extracted items to the number of items both correctly and erroneously extracted from the text; noise is the ratio of the faulty extracted items to all the achieved extractions. Recall is the ratio of correctly extracted items to the number of items actually present in the text. The problem consists in improving both precision and recall.

\subsection{Recall improvement}

A usual technique to improve the recall consists of enriching a text with a list of synonyms or near-synonyms for each word of that text. For example, all the synonyms of ``climb'' would be added to the document, even though some of those meanings have a remote semantic connection to the text. By this kind of enrichment, all the ways to express the same token (but not the same meaning) are taken into account.

This type of enrichment can be extended to synonymous expressions with a robust parser: syntactic dependencies and their arguments (the tokens belonging to the selected expression) are enlarged to dependencies that are generated out of the corresponding synonymous expressions.

The recall is usually optimised to the detriment of the precision with those techniques, since most words within a set of synonyms are themselves polysemous and are seldom equivalent for each of their meanings. Thus, a simply adding of all those polysemous synonyms in a document introduces meaning inconsistencies. Noise may stem from these inconsistencies.

\subsection{Reduction of noise -- Precision improvement}

We notice that improving the recall using synonyms may often increase the noise. Although identified in the domain of IE, this problem is not yet solved and it has a negative influence on the system effectiveness. Our purpose is to use the linguistic context of the polysemous tokens to identify their meanings and select contextual synonyms or synonymous expressions. This approach should improve the precision in comparison with adding all the synonyms.

\begin{figure}[h]
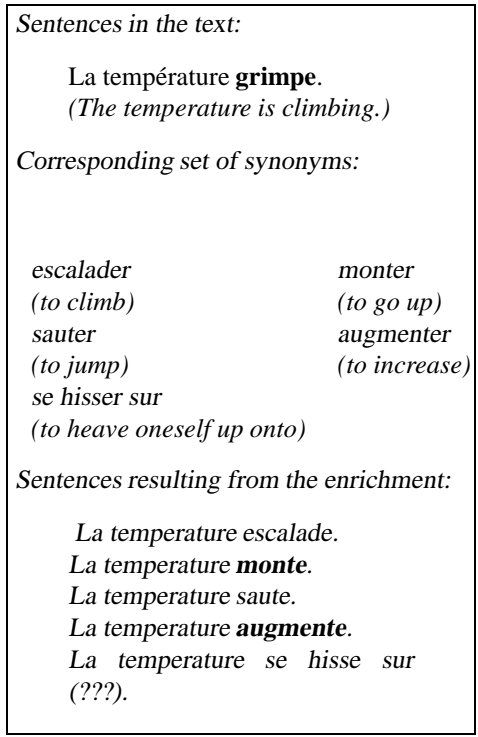

\begin{center}
\fbox{\parbox{6cm}{
\textsl{Sentences in the text:}
\begin{quote}
La temp\'erature \textbf{grimpe}.\\
\textit{(The temperature is climbing.)}
\end{quote}

\textsl{Corresponding set of synonyms:}\\
\begin{center}
\textsl{
\begin{tabular}{ll}
escalader   & monter\\
\textit{(to climb)}  & \textit{(to go up)}\\
sauter      & augmenter \\
\textit{(to jump)}   & \textit{(to increase)}\\
se hisser sur & \\
\textit{(to heave oneself up onto)} & \\
\end{tabular}}
\end{center}

\textsl{Sentences resulting from the enrichment:}
\begin{quote}
\textsl{
La temperature escalade.\\
La temperature \textbf{monte}.\\
La temperature saute.\\
La temperature \textbf{augmente}.\\
La temperature se hisse sur (???).
}\end{quote}
}}
\caption{Enrichment by a list of synonyms.}
\end{center}
\end{figure}

For example, the dictionary~\footnote{The dictionary we use is a French electronic one \cite{dubois-duboischarlier97}. We will give a more detailed information about it later.} entry for the word \textit{grimper} contains a set of 5 synonyms. If we use these synonyms to enrich the original text, we obtain five variations of the original sentence. Only the second and the fourth of the enriching variations are accurate in this context. The meteorological context associated with the word \textit{temp\'erature} in the dictionary should correctly discriminate the synonyms in this context: in the dictionary, each synonym of a lemma is associated with a meaning of this lemma and with the typical linguistic context of the lemma in this sense.

Consequently, we decided to use the linguistic context of the words that can be enriched to discriminate which synonyms should be used and which should not. The synonyms are stored in the dictionary according to the sense of each lemma. So, the task amounts to performing a lexical semantic disambiguation of the text and using synonymous expressions in the selected meanings to enrich the document.

\section{Enrichment method by WSD}

\subsection{Our experience in WSD}
\label{The WSD method}

We previously have developed a range of tools and techniques to perform Word Sense Disambiguation (WSD), for French and English. The basic idea  is to use a dictionary as a tagged corpus in order to extract semantic disambiguation rules, \cite{BRUN2002,BRUN2000,CBFS2000,DINI98,DINI99}. Since electronic dictionaries exist for many languages and they encode fine-grained reliable sense distinctions, be they monolingual or bilingual, we decided to take advantage of this detailed information in order to extract a semantic disambiguation rule database~\footnote{The English dictionary contained 39\,755 entries and 74\,858 senses, ie a polysemy of 1.88; the French dictionary contained 38\,944 entries and 69\,432 senses, ie a polysemy of 1.78}. 
 The disambiguation rules associate each word with a sense number taking the context into account. For bilingual dictionaries the sense number is associated with a translation, for monolingual dictionaries with a definition. WSD is therefore performed according to sense distinctions of a given dictionary.
The linguistic rules  have been created using functional dependencies provided by an incremental shallow parser (IFSP, \cite{Ait97}), semantic tags from an ontology (45 classes from WordNet \cite{FELDBAUM98} for English) as well as information encoded in SGML tags of dictionaries. This method comprises two stages, rule extraction and rule application.
\begin{itemize}
\item Rule extraction process: for each entry of the dictionary, and then for each sense of the entry, examples are parsed with the IFSP shallow parser. The shallow parsing task includes tokenization, morphological analysis, tagging, chunking, extraction of functional dependencies, such as subject and object (SUBJ(X, Y), DOBJ (X, Y)), etc. For instance, parsing the dictionary example attached to one particular sense {\bf S$_{i}$} of {\it drift} : 

{\it 1)The country is drifting towards recession.}

Gives as output the following chunks and dependencies :

[SC [NP The country NP]/SUBJ :v is drifting SC] [PP towards recession PP] 
SUBJ(country, drift)
VMODOBJ(drift, towards, recession)

Using both the output of the shallow parser and the sense numbering from the dictionary we extract the following semantic disambiguation rule:
When the ambiguous word ``drift'' has {\it country} as subject and/or {\it toward recession} as modifier, it can be disambiguated with its sense {\bf S$_{i}$}. We repeat this process as all dictionary example phrases in order to extract the word level rules, so called because they match the lexical context. 

Finally, for each rule already built, we use semantic classes from an ontology  in order to generalize the scope of the rules. In the above example the subject ``country'' is replaced in the semantic disambiguation rule by its ambiguity class. We call ambiguity class of a word, the set of WordNet tags associated with it. Each word level rule generates an associated class level rule, so called because it matches the semantic context: when the ambiguous word ``drift'' has a word belonging to the WordNet ambiguity class {\it noun.location} and {\it noun.group} as subject and/or a word belonging to the WordNet ambiguity class {\it noun.shape}, {\it noun.act}, and {\it noun.state} as modifier,  it disambiguates with its sense {\bf S$_{i}$}. Once all entries are processed, we can use the disambiguation rule database to disambiguates new unseen texts.
For French, semantic classes (69 distinctive characteristics) provided by the {\it AlethDic} dictionary \cite{Erli} have been used with the same methodology.

\item Rule application process: 
The rule applier matches rules of the semantic database against new unseen input text using a preference strategy in order to disambiguate words on the fly.
Suppose we want to disambiguate the word drift, in the sentence: 

{\it 2) In November 1938, after Kristallnacht, the world drifted towards military conflict.}

The dependencies extracted by the shallow parser, which might lead to a disambiguation, i.e., which involve \textit{drift}, are: 

SUBJ(world, drift)\\
VMODOBJ(drift, towards, conflict)

The next step tries to match these dependencies with one or more rules in the semantic disambiguation database. First, the system tries to match lexical rules, which are more precise. If there is no match, then the system tries the semantic rules, using a distance calculus between rules and semantic context of the word in the text~\footnote{The first parameter of this metric is the intersection of the rule classes and the context classes; the second one is the union of the rule classes and the context classes. Distance equals the ratio of intersection to union.}. In this particular case, the two rules previously extracted match the semantic context of {\it drift}, because {\it world} and {\it country} shares semantic classes according to WordNet, as well as {\it conflict} and {\it recession}.
\end{itemize}
 The methodology attempts to avoid the data acquisition bottleneck observed in WSD techniques. Thanks to this methodology, we built all-words (within the limits of the used dictionary) unsupervised Word Sense Disambiguator for French (precision: 65\%, recall: 35\%) and English (precision: 79\%, recall: 34\%).

\subsection{Xerox Incremental Parser (XIP)}

IFSP, which was used in the first experiments on semantic disambiguation at Xerox, has been implemented with transducers. Transducers proved to be an interesting formalism to implement quickly an efficient dependency parser, as long as syntactic rules would only be based on POS. The difficulty of using more refined information, such as syntactic features, drove us to implement a specific platform that would keep the same strategies of parsing as in IFSP, but would no longer rely on transducers.

This new platform \cite{AitAl01,Roux99} comprises different sorts of rules that chunk and extract dependencies from a sequence of linguistics tokens, which is usually but not necessarily a sentence. The grammar of French that has been developed computes a large number of dependencies such as {\it Subject, Object, Oblique, NN} etc. These dependencies are used in specific rules, the disambiguation rules, to detect the syntactic and semantic information surrounding a given word in order to yield a list of words that are synonyms according to that context. Thus, a disambiguation rule manipulates together a list of semantic features originating from dictionaries, and a list of dependencies that have been computed so far. The result is a list of contextual synonyms.

\noindent
If (Dependency$_0$(t, t$^0$) \& \dots \& Dependency$_n$(t,t$^k$) \& \dots\ attribute$_p$(t$^j$)=v$^u$) 

\emph{synonym}(t) = s$^0$,\dots,s$^n$.\\
where

t$^0$,\dots,t$^n$ is a list of token

s$^0$,\dots,s$^n$ a list of synonyms.

\begin{figure}[h]
\begin{center}
\fbox{\parbox{6cm}{
Example:
\begin{itemize}
\item La temp\'erature grimpe.\\ {\it (the temperature is climbing)}
\item La temp\'erature augmente.\\ {\it (the temperature is rising)}
\item L'alpiniste grimpe le mont Ventoux.\\ {\it (the alpinist climbs the mount
Ventoux)}
\item ???L'alpiniste augmente le mont Ventoux.\\ {\it (???the alpinist raises
the mount Ventoux)}
\end{itemize}
}}
\caption{Application of a disambiguation rule for enrichment.}
\end{center}
\end{figure}

The contextual synonymy between {\it grimper} and {\it augmenter} can be defined with the following rule. The feature \emph{MTO} is one of the semantic features that are associated with the entries of the Dubois dictionary. This feature is associated with each word that is connected to meteorology, such as {\it chaleur, froid, temp\'erature} (heat, cold, temp\'erature).

if (Subject({\it grimper}, X) AND feature(X, {\it domain})=MTO) synonym({\it
grimper}) = {\it augmenter}.

This rule applies on the above first example, {\it La temp\'erature grimpe}, but fails to apply on the third sentence, {\it L'alpiniste grimpe le mont Ventoux}, since the subject does not bear the MTO feature.

\subsection{Which WSD for which enrichment?}

\subsubsection{A very rich dictionary information}

The new robust parser offers a flexible formalism and the possibility to handle semantic or other features. In addition to this parser, the semantic disambiguation now uses a monolingual French dictionary \cite{dubois-duboischarlier97}. This dictionary contains many kind of information in the lexical field as well as in the syntactic or the semantic one. From the 115\,229 entries of this dictionary, we can only use the 38\,965 ones that are covered by the morphological analyser. These entries represent 68\,588 senses, ie a polysemy of 1.76.

We build lexico-syntactic WSD rules using the methodology presented above (cf. section \ref{The WSD method}): examples of the dictionary are parsed; extracted syntactic relations and their arguments are used to create the rules. We also make the most of the domain indication (171 different domains) to generalize the example rules (see later for details) --  as previously done using WordNet for the English WSD and by AlethDic for the French one \cite{BRUN2002}.

We use the specificity of the dictionary to improve the disambiguation task as far as possible in order to maximize the enrichment of the documents. The information of this dictionary is divided into several fields: domain, example, morphological variations, derived or root words, synonyms, POS, meaning, estimate of use frequency in the common language; in the verbal part of the dictionary only, syntactico-semantic class and subcategorization patterns of the arguments of the verb. Resulting WSD rules are spread over three levels reflecting the abstraction register of the dictionary fields.

\subsubsection{Disambiguation rules at various levels}

We build a disambiguation rule database at three levels: rules at word level (23\,986), rules at domain level (22\,790) and rules at syntactico-semantic level (40\,736).

Word level rules use lexical information from the examples. They correspond to the basic rules in the previous system, which use constraints on words and syntactic relations. These dependencies are extracted from the illustrative examples from the dictionary.

\begin{figure}[h]
\begin{center}
\fbox{\parbox{6cm}
{
L'avion de la soci\'et\'e \textbf{d\'ecrit} un large cercle avant de (\dots)\\
\textit{(The company's plane \textbf{describes} a wide circle before (\dots))}\\
SUBJECT(d\'ecrire,avion)\\
OBJECT(d\'ecrire,cercle)\\
\\
\textsl{Example in the dictionary for the entry ``d\'ecrire'':}\\
L'avion d\'ecrit un cercle.\\
\textit{(The plane describes a circle.)}\\
SUBJECT(d\'ecrire,avion)\\
OBJECT(d\'ecrire,cercle)
}}
\caption{WSD at word level.}
\end{center}
\end{figure}

Rules at domain level are generalized from word level rules: instead of using the words of the examples as arguments of the syntactic relations in the rules, we replace them by the domains they belong to. These rules correspond to the class level rules in the previous system, but an improvement in comparison with them is that in some cases, we can discriminate the right domain if the argument is polysemous. This is mainly due to the internal consistency of the dictionary that enables the correspondences of domain across different arguments of a dependency. The consistency should help to reduce the noise.

\begin{figure}[h]
\begin{center}
\fbox{\parbox{6cm}{
L'escadrille d\'ecrit son approche vers l'a\'eroport o\`u (\dots)\\
\textit{(The squadron describes its approach to the airport where (\dots))}\\
SUBJECT(d\'ecrire,escadrille$[$dom:AER$]$)\\
OBJECT(d\'ecrire,approche$[$dom:LOC$]$)\\
\\
\textsl{Example in the dictionary for the entry ``d\'ecrire'':}\\
L'avion d\'ecrit un cercle.\\
\textit{(The plane describes a circle.)}\\
SUBJECT(d\'ecrire,avion$[$dom:AER$]$)\\
OBJECT(d\'ecrire,cercle$[$dom:LOC$]$)
}}
\caption{WSD at domain level.}
\end{center}
\end{figure}

We don't rule out the possibility of using other lexico-semantic resources to generalize or expand this kind of rules, as we did previously using French EuroWordNet or AlethDic. These lexicons present the advantage of a hierarchical structure that doesn't exist for the domain field in the Dubois dictionary. Nevertheless, we will encounter the problem of the mapping of the various resources used by the system to avoid inconsistencies between them, as shown in \cite{IdeVeronis90,LuxAl98,BRUN2002}.

The third level of the rules currently in use in the semantic disambiguator is the syntactico-semantic one. The abstraction level of these rules is even higher than in the domain level. They are built from a syntactic pattern of subcategorization that indicates the typical syntactic construction of the current entry in its current meaning. Although the distinction between the arguments is very general -- they are differentiated from human, animal and inanimate -- our examination of the verbal dictionary indicates that, for 30\% of the polysemous entries, this kind of rules is sufficient to choose the appropriate meaning.

\begin{figure}[h]
\begin{center}
\fbox{\parbox{6cm}{
L'escadrille d\'ecrit son approche vers l'a\'eroport where (\dots)\\
\textit{(The squadron describes its approach to the airport where(\dots))}\\
SUBJECT(d\'ecrire,escadrille$[$dom:AER$]$)\\
OBJECT(d\'ecrire,approche$[$dom:LOC$]$)\\
\\
\textsl{Subcategorisation for the entry ``d\'ecrire'':}\\
Transitive verb;\\
Subject inanimate.\\
SUBJECT(d\'ecrire,?$[$subcat:inanimate$]$) \& OBJECT(d\'ecrire,?)
}}
\caption{WSD at lexico-semantic level.}
\end{center}
\end{figure}

\subsection{Enrichment at various levels}

WSD is not an end in itself. In our system, it is a means to select appropriate information in the dictionary to enrich a document. The quality and the variety of this enrichment vary according to the quality and the richness of the information in the dictionary. The variety of information allows several kind of enrichment.

For the specific task of information extraction, an index of the documents whose information is likely to be extracted is built. It allows the classification of all the linguistic realities extracted from text analysis. These realities are listed according to the XIP-formalism: syntactic relations, arguments, and features attached to the arguments. The enrichment is done inside the index because dependencies can be added without affecting the original document.

\subsubsection{Lexical level}

Replacing a word by its contextual synonyms is the easiest way to perform enrichment. This method of recall improvement is very common in IE, but in our system, the enrichment is targeted according to the context thanks to the semantic disambiguation. This process often reduces the noise. The enrichment is achieved by copying the dependencies containing the disambiguated word and by replacing this word by one of its synonyms.

\begin{figure}[h]
\begin{center}
\fbox{\parbox{6cm}{
La temp\'erature grimpe.\\
\textit{(The temperature is climbing.)}\\
\\
\textsl{Original index:}\\
SUBJECT(grimper,temp\'erature)\\
\\
\textsl{Set of targeted synonyms:}\\
monter, augmenter.\\
\\
\textsl{Enriched index:}\\
SUBJECT(grimper,temp\'erature)\\
\textbf{SUBJECT(monter,temp\'erature)\\
SUBJECT(augmenter,temp\'erature)}
}}
\caption{Enrichment at lexical level.}
\end{center}
\end{figure}

\subsubsection{Lexico-syntactic level}

The lexico-syntactic level of enrichment is more complex to achieve. The task consists in replacing a word by a multi-word expression (more than 14\,000 synonyms are multi-word expressions in our dictionary) or in replacing a multi-word expression by a word, taking into account the words (lexical) and the dependencies between them (syntactic):
\begin{itemize}
\item Replacing a word by a multi-word expression (see figure \ref{LSL}): 
  \begin{itemize}
    \item Parse the multi-word expression to obtain dependencies;
    \item Match the corresponding dependencies in the text;
    \item Instantiate the missing arguments with the text arguments.
  \end{itemize}
\item Replacing a multi-word expression by a word:
  \begin{itemize}
    \item Identify the POS of the word;
    \item Select dependencies implying one and only one word of the multi-word expression;
    \item Eliminate dependency where this word has a different POS;
    \item Replace this word with its synonym in the remaining dependencies.
  \end{itemize}
\end{itemize}

\begin{figure}[h]
\begin{center}
\fbox{\parbox{6cm}{
Le sp\'ecialiste a \'edit\'e un manuscrit tr\`es ab\^\i m\'e.\\
\textit{(The specialist published a very damaged manuscript.)}\\
\\
\textsl{Original index:}\\
SUBJECT(\'editer,sp\'ecialiste)\\
OBJECT(\'editer,manuscrit)\\
\\
\textsl{Targeted synonymous expression:}\\
\'etablir l'\'edition critique de\\
\\
\textsl{Extracted dependencies from the expression:}\\
SUBJECT(\'etablir,?)\\
OBJECT(\'etablir,\'edition)\\
EPITHET(\'edition,critique)\\
PP(\'edition,de,?)\\
\\
\textsl{Enriched index:}\\
SUBJECT(\'editer,sp\'ecialiste)\\
OBJECT(\'editer,manuscrit)\\
\textbf{SUBJECT(\'etablir,sp\'ecialiste)\\
OBJECT(\'etablir,\'edition)\\
EPITHET(\'edition,critique)\\
\textit{PP(\'edition,de,manuscrit)}}
}}
\caption{Enrichment at lexico-syntactic level.}
\label{LSL}
\end{center}
\end{figure}

Since our work is based on the Dubois dictionary -- whose entries are single words -- most of the enrichment is one-to-one word. When a multi-word expression appears in the synonyms list, a single word has to be replaced by a multi-word expression, and the inverse process can be achieved if necessary. The complex case of replacing a multi-word expression by another multi-word expression could arise, but we never encounter this situation. The replacement of a multi-word expression by another is not yet implemented because of the complexity of the process. Nevertheless, the system relies on relations and arguments that are easy to handle, very simple and modular. These characteristics should allow us to bypass the inherent complexity of these structures.

\subsubsection{A semantic level example}

Syntactico-semantic fields in the dictionary allow a third enrichment level. The syntactico-semantic class structure contains very useful information that makes it possible to link verbs that are semantically related but lexically and syntactically very different. It might be interesting to semantically link \textit{vendre} (``to sell'', class D2a) and \textit{acheter} (``to buy'', class D2c) even though their respective actors are inverted. For example, \textit{le marchand vend un produit au client} (the trader sells a product to the customer) bears the same meaning as \textit{le client ach\`ete un produit au marchand} (the customer buys a product from the trader). The semantic class gives a general meaning of the verb(D2, meaning \textit{donner, obtenir}, to give, to obtain), while the syntactic pattern (a for \textit{vendre}: \textit{fournir qc à qn}, to supply so with sth, transitive with a \textit{à} oblique compliment, c for \textit{acheter}: \textit{prendre qc à qn}, to take sth to so, transitive with a \textit{à} oblique compliment)  yields the semantic realization.

\begin{figure}[h]
\begin{center}
\fbox{\parbox{6cm}{
Le papa offre un cadeau \`a sa fille.\\
\textit{(The father is giving a present to his daughter.)}\\
\\
\textsl{Original index:}\\
SUBJECT(offrir,papa)\\
OBJECT(offrir,cadeau)\\
OBLIQUE(offrir,fille)\\
\\
offrir 01: D2a (to give sth to sb)\\
D2a corresponds to D2e (receive, obtain sth from sb).\\
recevoir 01: D2e\\
\\
\textsl{Enriched index:}\\
SUBJECT(offrir,papa)\\
OBJECT(offrir,cadeau)\\
OBLIQUE(offrir,fille)\\
\textbf{SUBJECT(recevoir,fille)\\
OBJECT(recevoir,cadeau)\\
????(recevoir,de,papa)}
}}
\caption{Enrichment at semantic level.}
\end{center}
\end{figure}

In a same perspective, a syntactico-semantic class constitutes another synonym set. Since this set is too general and too imprecise, it cannot be used to enrich a document. Still, it can be used as a last resort to enrich the query side when other methods have failed. We will not use this set as enrichment, but only to match a query by the class if the enrichment fails.

\section{Evaluation}

Though the method presented in this article is based on previous works, the use of other tools and lexical resource may have extended the potential of WSD rules. In particular, it is possible that the number of domains increase precision, and the use of subcategorization patterns may ensure more general rules to increase recall.

The partial evaluation we performed concerns 604 disambiguations in a corpus of 82 sentences from the French newspaper \textit{Le Monde}. Precision in WSD is ratio of correct disambiguations to all disambiguations performed; recall is ratio of correct disambiguations to all possible disambiguations in the corpus. We distinguish the mistakes due to the method and the ones linked to our analysis tools in order to identify what we have to improve in order to increase the performance. These results are promising since both precision and recall are better than in the previous system.

\begin{table}[h]
\begin{center}
\begin{tabular}{|l|r|r|}
\hline
Tokenization mistakes  & 44           & 7.28\%\\
\hline
Tagging mistakes       & 19           & 3.15\%\\
\hline
Parsing mistakes       &  9           & 1.49\%\\
\hline
WSD mistakes           & 84           & 13.91\%\\
\hline
\textbf{Precision}     & \textbf{448} & \textbf{74.17\%}\\
\hline
\textbf{Recall}        &              & \textbf{43.61}\%\\
\hline
\end{tabular}
\caption{WSD method evaluation.}
\end{center}
\end{table}

We note some remarks about this evaluation:
\begin{enumerate}
\item The lexicon used to perform tokenization has been modified in order to include additional information from the dictionary. We noticed during this evaluation some problems of coverage;
\item For this first prototype, we do not yet establish a strategy for cases in which multiple rules match. If more than one rule can be applied to the context, the sense is randomly chosen among the ones suggested by the matching rules~\footnote{This random choice is only performed for this evaluation and not in a IE perspective, since noise is better than silence in this field.};
\item Conversely, we do not yet try a strategy using the domain of disambiguated words as a general context to choose the corresponding meaning of a word to disambigate.
\end{enumerate}

During the evaluation, we also notice that when a result was correct, the suggested synonymous expressions were always correct for the disambiguated word in this context. Our method for an optimized enrichment is validated.

\section{Conclusion}

In this paper, we present an original method for processing documents, preparing the text for information extraction. The goal of this processing is to expand each concept by the largest list of contextualy synonymous expressions in order to match a request corresponding to this concept.

Therefore, we implement an enrichment methodology applied to words and multi-word expressions. In order to perform the enrichment task, we have decided to use WSD to contextually identify the appropriate meaning of the expressions to expand. Inconsistent enrichment by synonyms is currently known as a major cause of noise in Information Extraction systems.  Our strategy lets the system target the enriching synonymous expressions according to the semantic context. Moreover, this enrichment is achieved  not only with single synonymous words, but also with multi-word expressions that might be more complex than simple synonyms.

The WSD task and the resulting enrichment stage are achieved using syntactic dependencies extracted by a robust parser: the WSD is performed using lexico-semantic rules that indicate the preferred meaning according to the context. The linguistic information extracted from the analysis of the documents is indexed for the IE task. This index also stores additional new dependencies stemming from the enrichment process.

The utilization of a unique, all-purpose dictionary to achieve WSD and enrichment ensures the consistency of the methodology. Nevertheless, the information quality and richness of the dictionary might determine the system effectiveness.

The evaluation validates the quality of our method, which allows a great deal of lexical enrichment with less noise than is introduced by other enrichment methods. We have also indicated some ways our method could be expanded and our analysis tools could be improved. Our next step will be to test the effect of the enrichment in an IE task.

The method is designed to achieve a generic IE task, and the tools and resources are developed to process text data at a lexical level as well as at a syntactic or semantic level.


\bibliographystyle{lrec2000}
\bibliography{lrec2002} 

\begin{thebibliography}{}

\bibitem[\protect\citename{A\"it-Mokhtar and Chanod}1997]{Ait97}
Salah A\"it-Mokhtar and Jean-Pierre Chanod.
\newblock 1997.
\newblock Subject and object dependecy extraction using finite-state
  transducers.
\newblock In {\em ACL'97 Workshop on Information Extraction and the Building of
  Lexical Semantic Resources for NLP Applications}, pages 71--77, 7-12 juillet.

\bibitem[\protect\citename{A\"it-Mokhtar \bgroup et al.\egroup }2001]{AitAl01}
Salah A\"it-Mokhtar, Jean-Pierre Chanod, and Claude Roux.
\newblock 2001.
\newblock A multi-input dependency parser.
\newblock In {\em Proceedings of the Seventh International Workshop on Parsing
  Technologies}, pages 201--204, Beijing, China, 17-19 October. IWPT-2001,
  Tsinghua University Press.

\bibitem[\protect\citename{Brun and Segond}2001]{CBFS2000}
Caroline Brun and Fr\'ed\'erique Segond.
\newblock 2001.
\newblock Semantic encoding of electronic documents.
\newblock {\em International Journal of Corpus Linguistics}.

\bibitem[\protect\citename{Brun \bgroup et al.\egroup }2001]{BRUN2002}
Caroline Brun, Bernard Jacquemin, and Fr\'ed\'erique Segond.
\newblock 2001.
\newblock Exploitation de dictionnaires \'electroniques pour la
  d\'esambiguïsation s\'emantique lexicale.
\newblock {\em Traitement Automatique des Langues}, 42(3):667--690.

\bibitem[\protect\citename{Brun}2000]{BRUN2000}
Caroline Brun.
\newblock 2000.
\newblock A client/server architecture for word sense disambiguation.
\newblock In {\em Proceedings of Coling'2000}, pages 132--138, Saarbrücken,
  Deutschland.

\bibitem[\protect\citename{Dini \bgroup et al.\egroup }1998]{DINI98}
Luca Dini, Vittorio Di~Tomaso, and Fr\'ed\'erique Segond.
\newblock 1998.
\newblock Error driven word sense disambiguation.
\newblock In {\em Proceedings of the Conference COLING-ACL'98}, pages 320--324,
  Montr\'eal, août. COLING-ACL.

\bibitem[\protect\citename{Dini \bgroup et al.\egroup }2000]{DINI99}
Luca Dini, Vittorio Di~Tomaso, and Fr\'ed\'erique Segond.
\newblock 2000.
\newblock Ginger ii: an example-driven word sense disambiguator.
\newblock {\em Computer and the Humanities. Special Issue on SENSEVAL},
  34(1-2):121--126, avril.

\bibitem[\protect\citename{Dubois and
  Dubois-Charlier}1997]{dubois-duboischarlier97}
Jean Dubois and Fran\c~coise Dubois-Charlier.
\newblock 1997.
\newblock {\em Dictionnaire des verbes fran\c cais}.
\newblock Larousse, Paris.
\newblock This dictionary exists in an electronic version and is accompanied by
  the corresponding electronic \textit{Dictionnaire des mots fran\c cais.}

\bibitem[\protect\citename{Fellbaum}1998]{FELDBAUM98}
Christiane Fellbaum, 1998.
\newblock {\em WordNet: an electronic lexical database}, chapter Semantic
  Network of English Verbs, pages 69--104.
\newblock The MIT Press, Cambridge, Massachusetts.

\bibitem[\protect\citename{Gsi}1993]{Erli}
Gsi-Erli, France, 1993.
\newblock {\em Le dictionnaire AlethDic}, 1.5 edition, Mars.

\bibitem[\protect\citename{Ide and V\'eronis}1990]{IdeVeronis90}
Nancy Ide and Jean V\'eronis.
\newblock 1990.
\newblock Mapping dictionaries: A spreading activation approach.
\newblock In {\em Proceedings of the 6th Annual Conference of the Centre for
  the New Oxford English Dictionary}, pages 52--64, Waterloo, Ontario.

\bibitem[\protect\citename{Lehnert}1990]{Lehnert90}
Wendy Lehnert.
\newblock 1990.
\newblock Symbolic/subsymbolic sentence analysis: Exploiting the best of two
  worlds.
\newblock In J.~Barnden and J.~Pollack, editors, {\em Advances in Connexionist
  and Natural Computation Theory}, volume~1, pages 135--164. Ablex Publishers,
  Norwood, NJ.

\bibitem[\protect\citename{Lux \bgroup et al.\egroup }1999]{LuxAl98}
Veronika Lux, Corinne Jean, and Fr\'ed\'erique Segond.
\newblock 1999.
\newblock Wsd evaluation and the looking glass.
\newblock In {\em Proceedings of TALN-99}, Cargese. TALN-99.

\bibitem[\protect\citename{Riloff and Lorenzen}1999]{RiloffLorenzen99}
Ellen Riloff and Jeffrey Lorenzen.
\newblock 1999.
\newblock Extraction-based text categorization: generating domain-specific role
  relationships automatically.
\newblock In T.~Strzalkowski, editor, {\em Natural Language Information
  Retrieval}. Kluwer Academic Publisher.

\bibitem[\protect\citename{Riloff}1993]{Riloff93}
Ellen Riloff.
\newblock 1993.
\newblock Automatically constructing a dictionary for information extracting
  tasks.
\newblock In {\em Proceedings of the Eleventh National Conference on Artificial
  Intelligence}, pages 811--816. AAAI Press / MIT Press.

\bibitem[\protect\citename{Roux}1999]{Roux99}
Claude Roux.
\newblock 1999.
\newblock Phrase-driven parser.
\newblock In {\em Proceedings of VEXTAL'99}, pages 235--240, Venezia, Italia.
  VEXTAL'99.

\bibitem[\protect\citename{Soderland \bgroup et al.\egroup
  }1995]{SoderlandAl95}
Stephen Soderland, David Fisher, Jonathan Aseltine, and Wendy Lehnert.
\newblock 1995.
\newblock Crystal: Inducing a conceptual dictionary.
\newblock In {\em Proceedings of the Fourteenth International Joint Conference
  on Artificial Intelligence}, pages 1314--1320. IJCAI-95.

\bibitem[\protect\citename{Soderland}1999]{Soderland99}
Stephen Soderland.
\newblock 1999.
\newblock Learning information extraction rules for semi-structured and free
  text.
\newblock {\em Machine Learning}, 34(1):233--272.

\end{thebibliography}

\end{document}